\documentclass[10pt,final,journal,twocolumn]{IEEEtran}
\ifCLASSINFOpdf
\else
\fi
\usepackage{cite}
\usepackage[cmex10]{amsmath}
\usepackage{amsthm}
\usepackage{algorithmic}
\usepackage{stfloats}
\usepackage{multicol,multienum}
\usepackage[dvips]{graphicx}
\graphicspath{{./figurespdf/}}
\usepackage{algorithm} 
\usepackage{algorithmic} 
\usepackage{multirow} 
\usepackage{amsmath}
\usepackage{xcolor}
\usepackage{subfigure}

\usepackage{amssymb}
\usepackage{color}

\begin{document}

\title{Self-Organizing Relay Selection in UAV Communication Networks: A Matching Game Perspective}

\author{Dianxiong~Liu,
        Yuhua~Xu,
        Jinlong~Wang,
        Yitao~Xu,
        Alagan Anpalagan,
        Qihui~Wu,
        Hai~Wang,
        and~Liang~Shen

\thanks{This work was supported by the National Natural Science Foundation of China under Grant No. 61771488, No. 61671473 and No. 61631020, in part by the Natural Science Foundation for Distinguished Young Scholars of Jiangsu Province under Grant No. BK20160034, and in part by the Open Research Foundation of Science and Technology on Communication Networks Laboratory.}
\thanks{Dianxiong Liu, Yuhua Xu, Jinlong Wang, Yitao Xu, Hai Wang and Liang~Shen are at the College of Communications Engineering, Army Engineering University, Nanjing, China (e-mails: dianxiongliu@163.com, yuhuaenator@gmail.com, wjl543@sina.com, yitaoxu@126.com, haiwang@ieee.org, shenliang671104@sina.com).}
\thanks{Alagan Anpalagan is with the Department of Electrical and Computer Engineering, Ryerson University, Toronto, Canada (e-mail: alagan@ee.ryerson.ca).}
\thanks{Qihui Wu is with the College of Electronic and Information Engineer, Nanjing University of Aeronautics and Astronautics, Nanjing, China (e-mail: wuqihui2014@sina.com).   }
}

\IEEEpeerreviewmaketitle
\maketitle

\begin{abstract}
For large unmanned aerial vehicle (UAV) networks, the timely communication is needed to accomplish a series of missions accurately and effectively. The relay technology will play an important role in UAV networks by helping drones communicating with long-distance drones, which solves the problem of the limited transmission power of drones. In this paper, the relay selection is seen as the entry point to improve the performance of self-organizing network with multiple optimizing factors. Different from the ground relay models, the relay selection in UAV communication networks presents new challenges, including heterogeneous, dynamic, dense and limited information characteristics. More effective schemes with distributed, fast, robust and scalable features are required to solve the optimizing problem. After discussing the challenges and requirements, we find that the matching game is suitable to model the complex relay model. The advantages of the matching game in self-organizing UAV communications are discussed. Moreover, we provide extensive applications of matching markets, and then propose a novel classification of matching game which focuses on the competitive relationship between players. Specifically, basic preliminary models are presented and some future research directions of matching game in UAV relay models are discussed.
\end{abstract}

\section{Introduction}
With the increasing development of the unmanned aerial vehicle (UAV) technology, the large-scale multi-UAV systems \cite{R2.6} play an important role in military and civilian fields, such as in wars, emergency communications and Internet of Things \cite{UAVcommunication}. Maintaining the communication of large-scale UAV networks becomes an important and timely issue.

From the perspective of requirement, the communication of UAVs can be classified into internal communication and external communication. In the external communication, UAV networks mainly exchange the information with the higher command center via the satellite communication. However, due to the limited transmission capacity, the heavy UAV-to-infrastructure communication hardware and the unreliable communication \cite{R2.11}, it is difficult for all UAVs to connect with the satellite. With large-scale deployment and intensive cooperation of UAVs, it is meaningful to discuss the internal communication \cite{R2.12} of UAVs. In more extreme cases, if satellite communications are disrupted, effective internal communications can guarantee the completion of the mission.

\begin{figure}
\centering
  \includegraphics[width=3.2in]{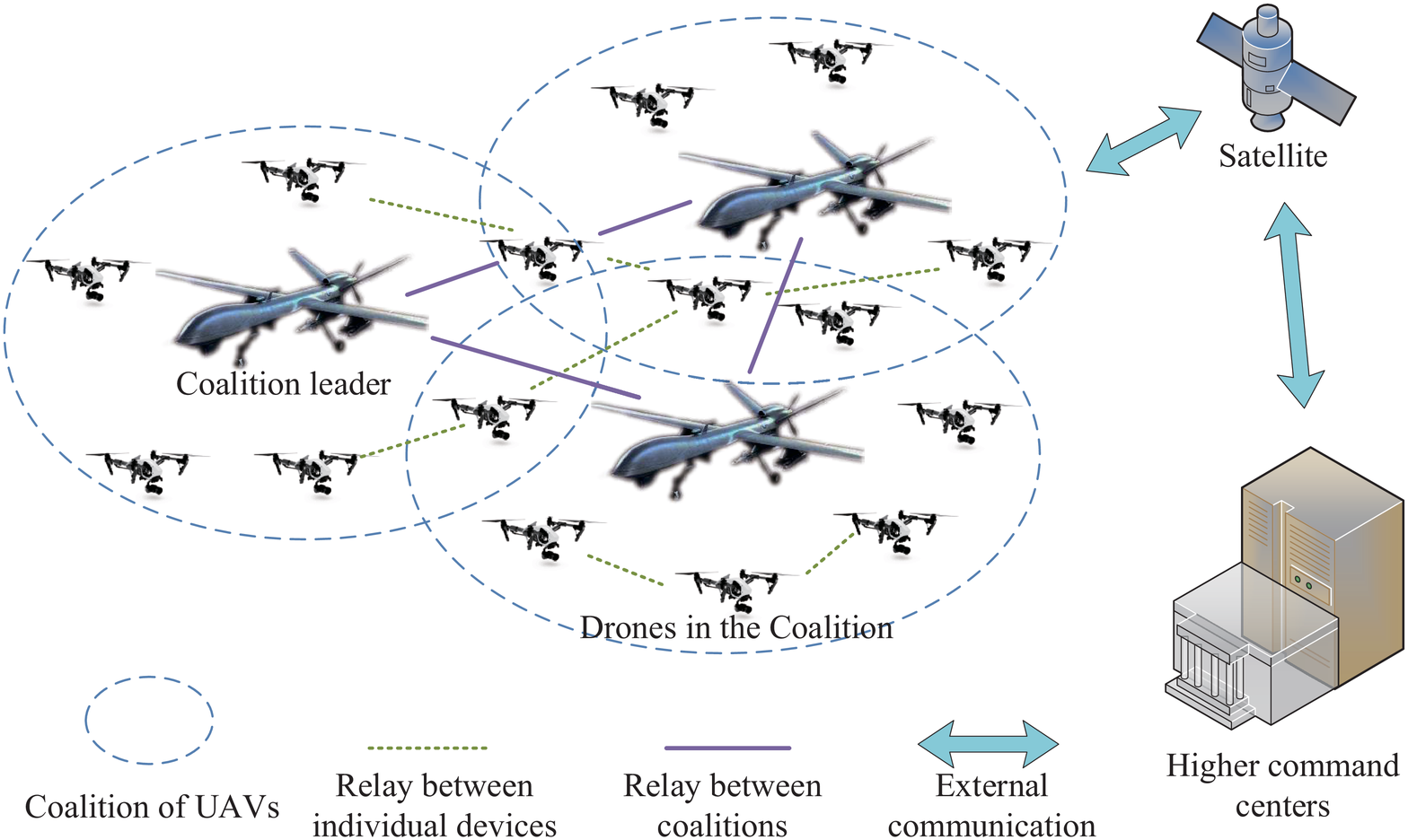}\\
  \caption{A model of the future UAV communication networks.}\label{network}
\end{figure}

In the internal communication, the technology of short haul communication is used. As shown in Fig. \ref{network}, due to the complexity of the spatial distribution, the UAV network needs to be divided into several coalitions to finish the comprehensive missions. Within the network, the internal communication of UAVs involves the control information exchange and the service information fusion. The information sharing mainly implements the information interaction among coalitions of UAVs, so as to configure the UAV coalitions and assign tasks. The information fusion mainly refers to the information exchange between the units within the UAVs, so drones can assess the overall situation and accomplish tasks.

Because of the transmission power constraint of drones, it is difficult to achieve reliable communications among the whole UAV network. Some drones should be used as relay devices to improve the quality of the communication. However, the large scale of UAVs, the self-organization of UAV coalition and the intensive external interference make the relay selection of UAV communication more difficult. Considering the complexity of the UAV communication, this article mainly provides a new perspective of developing distributed and robust relay selection technologies in UAV communication networks.

Compared with the additional trajectory planning of UAVs within the network, the relay transmission can optimize the UAV communication based on the formation configuration, without destroying the stability of the network. Moreover, based on relay transmission rather than additional mobility can save the energy consumption of UAVs so as to extend the endurance in the case of ensuring the normal communication of the UAV network. Therefore, this paper mainly develops the relaying optimization. Besides, dynamic strategies combined with the mobile UAVs are considered.

Due to the dynamic networks, limited abilities of drones and diversified tasks, it is hard to apply existing approaches of the relay selection to the communication scenarios of large scale UAVs alliance networks \cite{UAVcommunication}. Moreover, the problem of relay selection in UAV communication networks can not be solved in centralized methods, because lots of factors in relay selection strategies, such as power control, interference management and carrier selection, may result in heavy communication overhead. As a result, it is timely to develop distributed selection approaches for the future UAV relay networks.

After exploring some relay selection cases, this article analyzes some fundamental challenges and requirements of the self-organizing relay selection in UAV networks. Then, following the attractive features of matching game market \cite{matchinggame}, we propose and discuss some advantages of the matching game for self-organizing relay selection in UAV communication networks.

Matching game is powerful to tackle the problem of resource allocation by modeling the relationship of players between two distinct sets. It has recently attracted extensive attention in wireless networks \cite{futurematching,newmatchinggame}, such as cell association and cooperative spectrum sharing. Moreover, some preliminary matching solutions have been studied in ground relay models. However, the inherent features and fundamental challenges of relay selection in UAV communication networks need to be further studied. The matching game also needs to be developed to solve the resource assignment problems effectively. Compared with existing matching models \cite{futurematching,newmatchinggame}, i) we mainly focus on UAV relay models based on the featured matching models, the classification of relay models and special features of future UAV communication networks are explored; ii) focusing on the internal competition relationship among drones, we propose a novel classification of matching models in wireless networks, in which they are classified as matching with substitutability, with partial substitutability and without substitutability, respectively.

It is noted that from different perspectives of the multi-UAVs system, several UAV architectures were proposed in the existing work.

\begin{itemize}
  \item UAV swarms: The multi-UAVs system is analyzed from the perspective of the formation configuration, which mainly emphasizes the external physical features of UAV networks. Analysis from this perspective makes the models more intuitive, so problems of transportation planning and control \cite{R2.14} are generally solved by this point of view.
  \item Flying Ad-Hoc Network (FANET): It develops the networking technology of UAV networks. By analyzing the communication link, the architecture of multi-UAV communication networks can be optimized effectively. From this perspective, problems of the topology construction and routing protocol are mainly studied \cite{R2.12}.
  \item Multi-tier drone architecture: It is analyzed from the perspective of layers of structure \cite{R2.1}. By analyzing different types, flight altitudes and communication objectives of UAVs, multi-tiers UAVs are developed to improve the spectral efficiency of users in cellular networks. It is also an efficient perspective for the multi-UAVs system.
\end{itemize}

Different from these researches, we analyze the UAV networks from the perspective of resource optimization. The optimization of relay model among UAVs is discussed, which makes the topic more targeted so as to research into transmission situations, challenges, and corresponding solutions. We develop effective matching models of resource optimization based on the formation and communication architectures. The optimizing models proposed in this paper can be applied to the existing classification framework in different analytical perspectives.

The rest of this article is organized as follows. In Section II, after introducing the main classification of internal UAV relay models, the requirements and challenges of the relay selection in UAV communication networks are analyzed. In Section III, the matching game model in UAV networks is presented. In Section IV, extensive applications of matching game model in future UAV communication networks are discussed, new classification focusing on the relationship between players is proposed, and future research directions are given.

\begin{figure*}
\centering
  \includegraphics[width=5.2in]{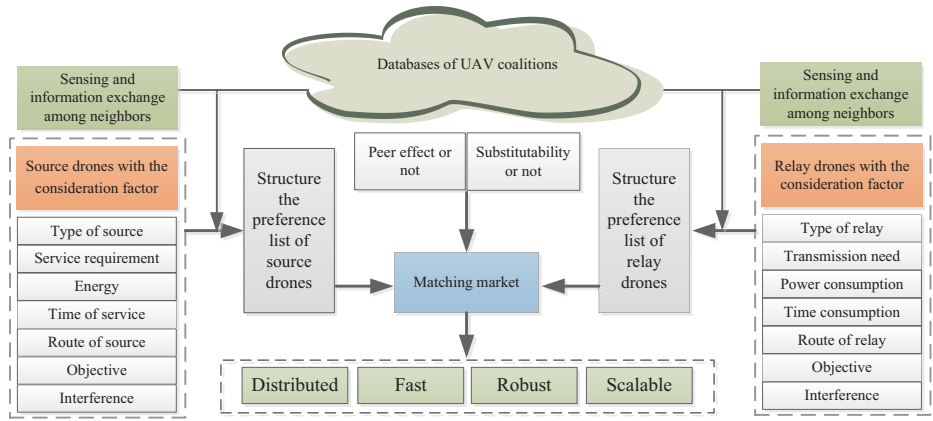}\\
  \caption{The structure of matching game model in UAV communication networks.}\label{matchingstructure}
\end{figure*}

\section{Relay selection for UAV communication networks}
In this section, we introduce the relay model from the perspective of the task planning. Moreover, by exploring the inherent features of UAV communication networks, we briefly discuss some fundamental challenges and requirements of optimizing relay selection in UAV communication networks.

\subsection{The classification of relay models in UAV communications}
Predictably, large UAV formations will be able to accomplish self-organizing air tasks, and a complicated task can be subdivided into several subtasks. Therefore, the UAV formations can be divided into different coalitions according to different tasks. In this case, the relay transmission will run through the whole UAV communication system. The main applications of relay selection in UAVs are shown as follows:
\subsubsection{Relay transmission between coalitions of UAVs}
Due to the different demands of subtasks, each coalition performs its subtasks and coordinates with each other so as to achieve the overall goal or complete the overall task. Therefore, the information exchange is necessary between the coalitions of UAV systems. Fig. \ref{network} shows that, one coalition of UAVs can communicate with remote coalitions relayed by other coalition leaders or drones which are located between two sides of the communication.
\subsubsection{Relay transmission between individual devices}
One drone needs to establish communication links with the coalition leader, and it also needs to communicate with the surrounding drones to coordinate the flight missions. As shown in Fig. \ref{network}, when one drone is too far away from the coalition leader to establish direct communication link, relay technology can be used to feed back the timely information. In addition, when drones need to share information with surrounding drones, they can also choose relay transmission to improve the transmission efficiency.

The main differences of the two types are given. Firstly, the coalition leader needs to communicate with other coalitions, which requires a strong transmission capability, while drones in the alliance can be small and reusable with lower transmission capability. According to the transmission capacity, full duplex and non-orthogonal multiple access technologies can be used in the inter-coalition transmission, while the half duplex with light loaded transmission can be used in the intra-coalition transmission. Moreover, in self-organizing systems, there is a competitive relationship between coalition leaders for channels access, while available resources within the coalition may be assigned by the coalition leader. In addition, when a drone needs to communicate with another drone in another coalition, the corresponding protocols of channel assignment may be required.
\subsection{Discussion of relay selection in UAV communication networks: Challenges and requirements}



The technical challenges of the relay selection optimization in UAV communication networks are discussed as follows:

\begin{itemize}\vspace{-0.1cm}
\item \textbf{Heterogeneous.} The ``heterogeneous" has two layers of meaning. On one hand, the network architecture is heterogeneous. The ``Gremlins" UAV program\footnote{Available: http://www.darpa.mil/news-events/2015-08-28} and multi-tier drone architecture \cite{R2.1} were proposed to develop the UAV system with various layers. There is a tight relationship between different levels of drones. For example, the reusable drones in ``Gremlins" have to maintain stable communications with large UAVs. On the other hand, in UAV networks, tasks are assigned by the system, and the communication is to ensure the completion of task, while the communication requirement of ground networks is generated spontaneously by users. Different from relay models in ground networks with single task transmission, drones may need to accomplish multiple tasks simultaneously.
\item \textbf{Dynamic networks.} The dynamic of UAV communications brings two aspects of challenges. The drastic change of the external environment and the rapid change of the formation may cause the previous communication link unavailable, which requires the rapid adjustment of the selection strategy of UAV networks. On the other hand, the dynamic change of the UAV formation may be the spontaneous movement to perform the task. The mobility of drones can be used to ferry the transmitted information, including the trajectory optimization \cite{R2.13} and the suitable selection of mobile UAVs.
\item \textbf{Dense deployment.} As mentioned in \cite{R2.6}, there may be hundreds of drones in the air simultaneously. The ground control is not reliable when the UAV network is out of the ground and carries out long-range missions on its own. How to optimize the self-organizing model in such a dense network has not been considered in the ground relay network. The resource optimization for dense deployment environment is different from which for sparse one. Particularly, the problems of transmission power constraint and interferences among drones are more serious.
\item \textbf{Limited information.} Without a powerful centralized controller and stable transmission conditions, UAVs with limited transmission power hardly obtain perfect environmental information. UAVs should make strategies according to limited and dynamic information from neighbor drones. In a heterogeneous, dynamic and distributed system, limited information as one constraint of frequency domain is inevitable.
\end{itemize}

By exploring the optimization features, we discuss some featured requirements of relay selection in UAV communication networks, which mainly include four features: distributed, fast, robust, and scalable.

\begin{itemize}\vspace{-0.1cm}
\item \textbf{Distributed:} In UAV communication networks, most drones are deployed randomly and dynamically in order to adapt to the flight tasks. The strategy of relay selection becomes more diversified due to dense deployment of drones. The centralized controller will process a large amount of data information and resource consumption. Thus, selection optimization problems for UAV communication networks are suitable to be solved in an effective self-organizing manner.
\item \textbf{Fast:} The dynamic feature of networks require drones spending shorter time to make decisions, because the location of drones may move and the sets of active communication drones are random. For example, one drone may change its flight path during the communications. Outdated strategies of relay selection may be meaningless in a short time, while the fast strategy can not only avoid the useless selection but also use the dynamic feature to improve the transmission.

\item \textbf{Robust:} The relay selection strategies should be robust for the dynamic environment. As discussed before, there are several complicated optimization factors of UAVs, such as dynamic communication tasks and varying flight paths. The obtained information may also be incomplete or corrupted by noise. Thus, the distributed selection solutions should be robust to address the problems of randomness, dynamics, and uncertainty in UAV communication networks.
\item \textbf{Scalable:} The resource optimization for dense deployment of UAV communication networks is different from that for sparse one. Thus, the self-organized selection schemes should have the ability to extend to the application of the dense UAV networks. Due to the limited transmission power of drones, multi-hop relay model will be applied in UAV networks, and the number of transmission hops is uncertain. The self-organizing relay selection models should be extended to multi-hop UAV communication networks.
\end{itemize}

\section{Matching market-based optimization for UAV communication models}
In this section, a brief introduction to the structure of matching game \cite{matchinggame,newmatchinggame,futurematching} is given. We will also discuss the special advantages of matching game in UAV communication networks.

In essence, the matching game is defined by two sets of players ($\mathcal{S},\mathcal{R}$) and two preference relations ${{\succ }_{i}},{{\succ }_{j}}$, permitting each player $i\in \mathcal{S},j\in \mathcal{R}$ constructing preference lists over one another, i.e., ranking the players in $\mathcal{S}$ and $\mathcal{R}$ respectively according to their individual preferences \cite{matchinggame}. Therefore, the matching game can be expressed as ${\cal G}\left( {{\cal S},{\cal R},{ \succ _i},{ \succ _j},{q_i},{q_j}} \right)$, where the maximum number of each player's ability to match is called quota, ${q}$. Each player uses the preference relation to rank the players in the opposing set. The proposed matching game can be fully represented once the preference of each player is defined.

\begin{figure}[t!]
\centering
\subfigure[A multi-hop relay is modeled as a multi-level matching market.]{
\label{multi-level} 
\includegraphics[width=3in]{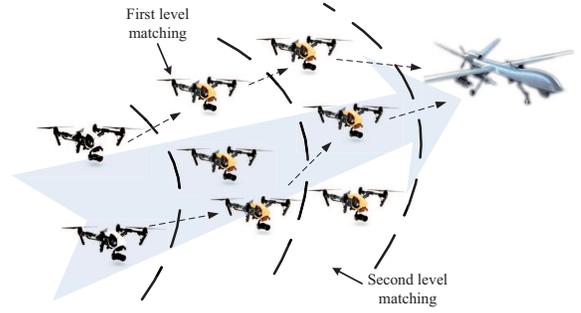}}
\centering
\subfigure[The matching model combines with the trajectory adjustment.]{
\label{dynamicmatching} 
\includegraphics[width=3in]{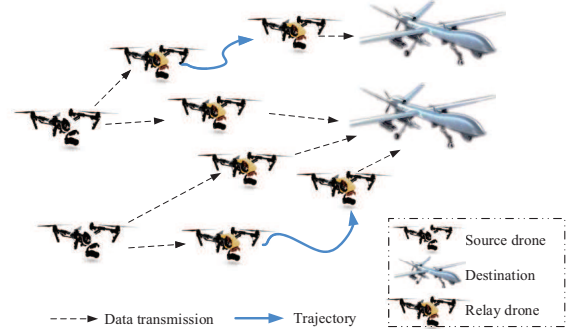}}
\caption{Examples of matching models in the future heterogeneous UAV networks.}
\label{figurelayerdynamic} 
\end{figure}

Fig. \ref{matchingstructure} shows the structure of matching model in UAV communication networks. Driven by different tasks, source drones select relay drones according to transmission objectives, types of transmission data, urgency of tasks, energy consumption and flight path. During the process, the relay selection and transmitting power can be adjusted according to the matching results. Similarly, relay drones will filter and adjust matching source drones according to several factors such as the preference of tasks, the energy consumption and the interference with other drones. It can be noted that the information required by source drones and relay drones is different, which is one of characteristics of the matching game model.

The information is obtained primarily from two ways: one is by sensing the environment and interacting with other drones, and the other is the information obtained from the coalition leader.

\begin{itemize}
  \item \textbf{The coalition leader as a temporary database:} Coalitions cooperate with each other to accomplish missions, and coalition leaders are used to summarize the information of drones within the coalition and then interact with the ground controller. Therefore, the coalition leader can be used as a temporary database to provide useful information for drones. For example, drones can request relative position and the plan of flight trajectory from the database. Compared to sensing the environment and interacting with other drones, the database approach is more efficient but not real-time.
  \item \textbf{Sensing and information exchange:} In the large-scale communication environment, drones cannot obtain the whole information from the coalition leader to make strategies, so they need to sense the surrounding environment information and exchange information among neighbors. For example, the channel occupancy can be obtained by energy detection or feature detection, and the information exchange can help with the task cooperation.
\end{itemize}

As discussed above, not all the information can be stored, drones need to make strategies by themselves to ensure more accurate optimization. Distributed transmissions have relatively strong robustness. When the database is unavailable, it is also possible for drones to make individual decisions by perceiving environment. Thus, it is necessary to develop the database-assisted distributed systems.

Based on different requirements and optimization factors, source drones construct individual matching lists \cite{matchinggame} in which the elements are relay drones. Similarly, relay drones choose source drones according to their own requirements. After constructing the preference list, the matching markets with or without peer effects \cite{futurematching} and substitutability are formed. The situation of substitutability will be discussed in Section IV. The matching game is good at modeling the relationship between source drones and relay drones.

Firstly, in distributed UAV models, source drones select the preferred relay drones according to their own transmission requirements. Meanwhile, relay drones filter source drones by their objectives and utilities. Matching game can define individual utilities for source drones and relay drones. The available algorithmic implementations allow a largely distributed solution to the problem of resource allocation without obtaining all of the information in networks.

Secondly, different from other game models, stable matching results are the primary objective in the matching market. In the distributed system, obtaining global optimum needs a long time to make decisions, while the fast changing network environment does not allow such kind of longer decision period. Pursuing the results of optimal relay selection in the whole network is not reasonable. The stable matching can respond to the changes effectively, and also maintain the stability of network communications. Therefore, stable matching is more suitable for UAV communication networks.

Thirdly, in the matching game, each player will learn and update the preference list during the matching process \cite{futurematching}. Due to the nature of the matching game, the previous learning list can guide the matching of players in the selection process. If current matching results are broken by the change of the environment in the matching process, players can make a faster adjustment according to the previous preference lists and the perception prediction results.

Finally, the extensibility of the matching scheme in various sizes of networks has been verified in the existing work such as \cite{multione}. Moreover, multi-tier matching models were preliminary designed and used in the wireless network \cite{R2.9}, which verified the availability of the matching game in the multi-level. In the future research, more organic combination can be developed. The discussion is given in Section IV. A.

\begin{figure*}
  \centering
  \includegraphics[width=5in]{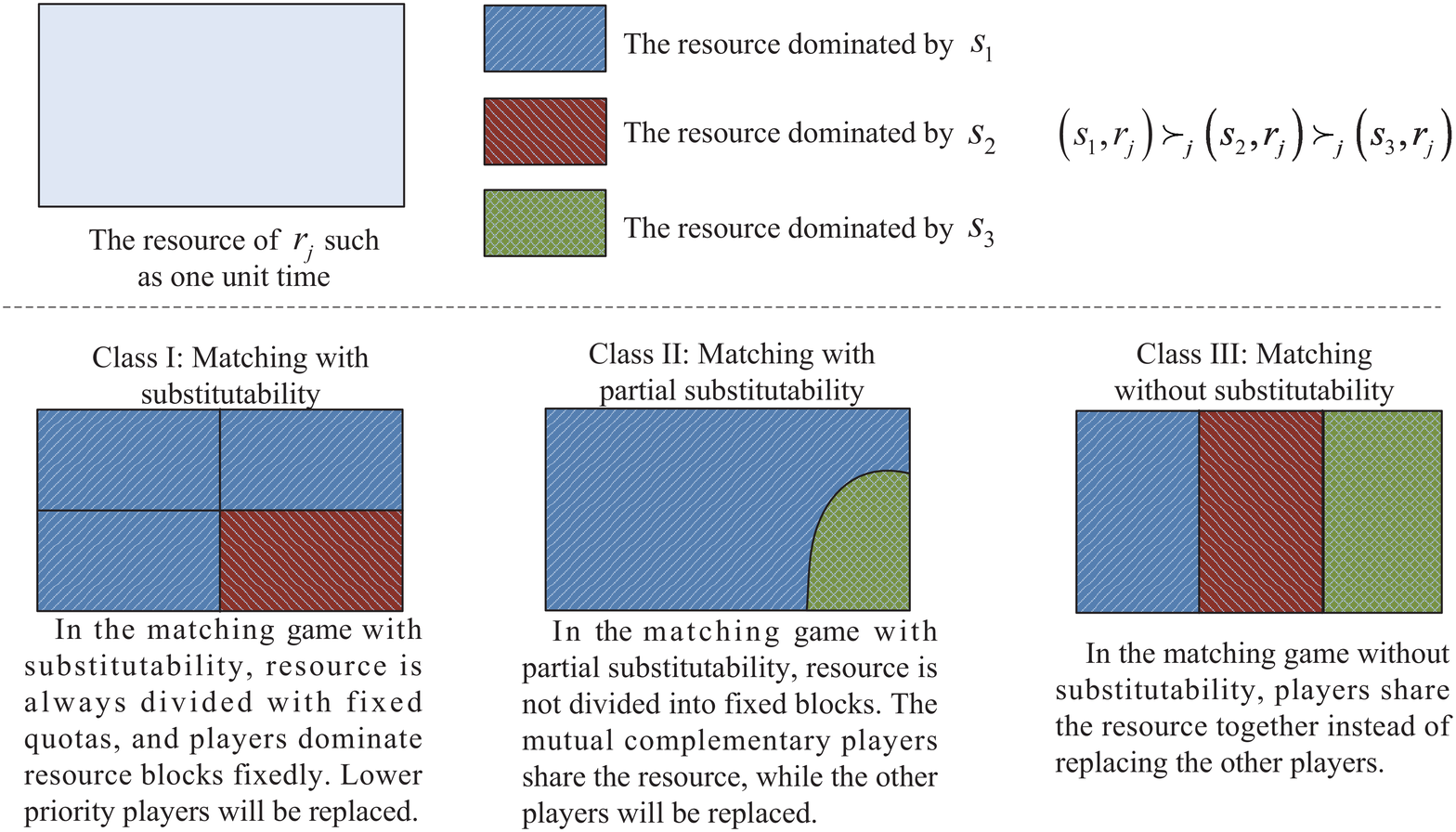}\\
  \caption{A novel classification of matching game focusing on the relationship of internal competition in wireless network. }\label{newclass}
\end{figure*}

\section{Matching market for future heterogeneous relay networks}
The authors in \cite{newmatchinggame,futurematching} discussed the classification of matching game according to the practical communication scenarios in ground wireless networks. The classification can cope with part of the problems of relay networks. However, novel features and applications of matching game in wireless networks can be further explored. Matching game theory need to be explored and extended so that new features for future relay models in UAV communication systems can be better applied.

\subsection{Extensive application of matching model}

\subsubsection{Multi-level matching game for multi-hop relay models}
In UAV communication networks with power constraint, source drones which cannot connect to destination drones by two-hop transmission should choose multi-hop relay models. Nowadays, distributed relay selection solution for multi-hop relay models have not been fully explored. Considering multiple optimization factors, it is necessary to model complex networks as suitable distributed models so as to optimize them with low complex methods. For the multi-hop communication, a multi-level matching market can be modeled, which is the extension from the general matching market. Global matching networks can be divided into multi-level markets and solved by matching approaches.

Fig. \ref{multi-level} shows that, if two source drones with limited transmission power want to connect with one destination drone, the communication should pass through three-hop transmissions. When source drones choose relay drones at the first level, the possible matching conditions of relay drones at the next level should be evaluated, which provides a good reference factor for the matching filter. Next, the selected relay drones will attend to the next level matching market and match relay drones at the next level. Because of the resource competition, source drones can adjust their selection strategies according to the matching results. Based on the multi-level matching model, the problem of multi-hop route selection can be solved by distributed methods. Therefore, multi-level matching provides an appropriate perspective for multi-hop relay models in UAV networks.

\subsubsection{Dynamic matching game combined with the trajectory optimization}

The trajectory and propulsion energy consumption are the key problems of UAV dynamics \cite{R1.27}, it is reasonable to develop the relative mobility of UAVs within networks. Different from the air-to-ground trajectory optimization \cite{R2.13, R1.27}, UAVs in the network have certain relative positions to finish the flight mission. The trajectory of multiple UAVs within the formation cannot be adjusted arbitrarily. During flying as a formation, drones can adjust their relative trajectories within the reasonable range. Therefore, it's a worthy direction to research into the cooperation transmission of multi-drones based on the topology adjustment and trajectory planning.

While UAVs are driven by tasks or impacted by the environment to adjust the relative position, they can be used to ferry information \cite{R2.13} from the other UAVs. Depending on the original trajectory instead of additional flight can save the additional propulsion energy as much as possible. As shown in Fig. \ref{dynamicmatching}, if UAVs adjust the flight formation during the communication, source UAVs can make use of the adjustment and select the appropriate communication mode effectively (select static UAVs or cooperate with mobile UAVs to help the transmission). Here, the dynamic matching model can be developed, where source drones and relay drones make strategies in advance according to the future dynamic adjustment. Different from the existing matching work which tried to avoid the influence of dynamics \cite{futurematching}, the adjustment of the UAV network can be prescient and used to improve the quality of transmission.

\subsubsection{Matching game with and without substitutability}
Conventional classification consists of one-to-one model, many-to-one model and many-to-many model, which is one of the classifications based on the structure of matching market. Authors in \cite{futurematching} classified the matching market as classic matching, matching with externalities and matching with dynamics, which do not refer to the inherently competitive relationships between players. To capture the features of inherent relationship of players in the matching game, beyond the existing classes, the classification according to the relationship of competitors is discussed following and illustrated in Fig. \ref{newclass}.

\begin{figure*}[t!]
\centering
\subfigure[The global satisfaction of the matching game with partial substitutability in relay models with heterogeneous tasks.]{
\label{performancecompared10relays1000times} 
\includegraphics[width=2.3in]{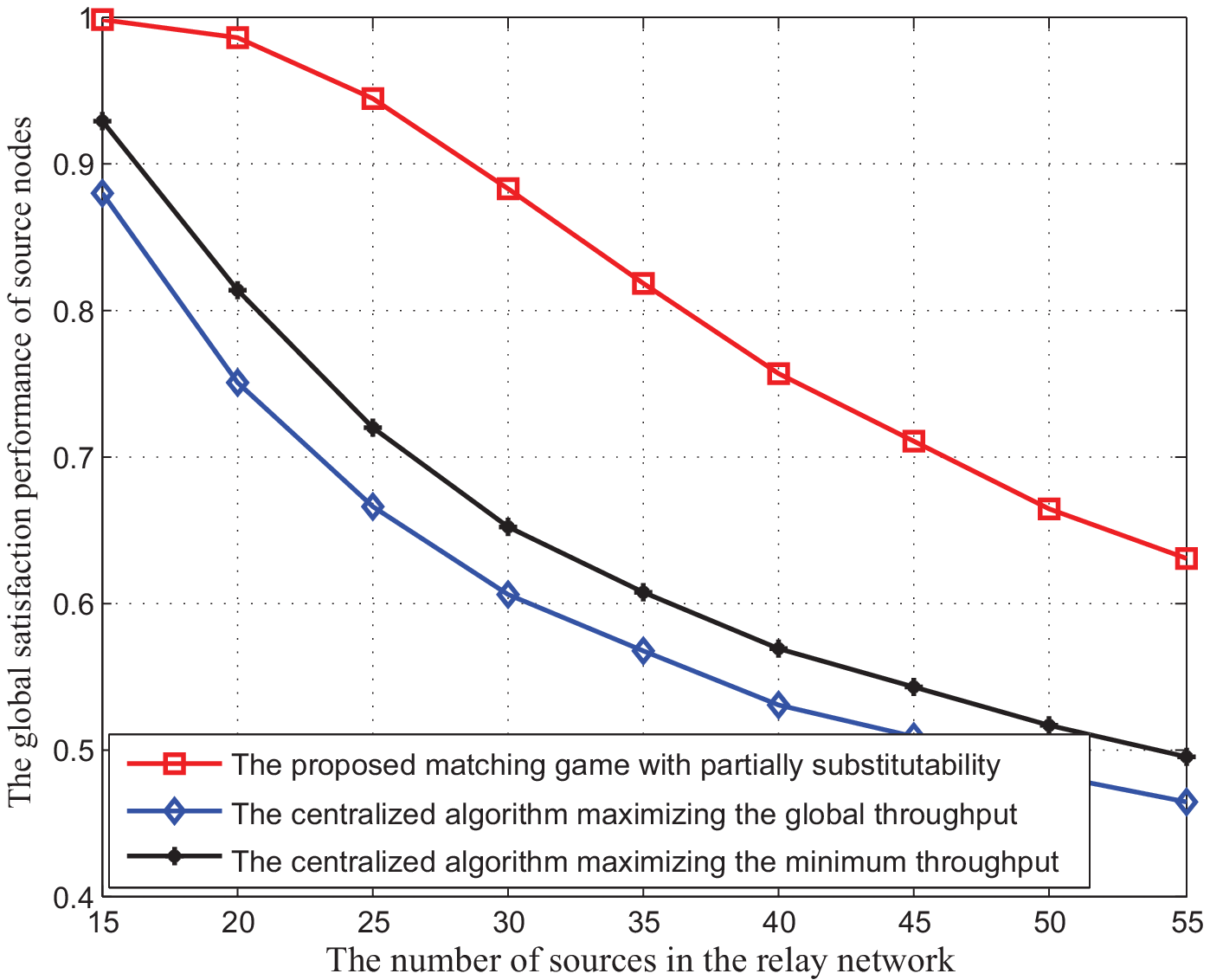}}
\centering
\subfigure[The matching game without substitutability achieving the nearly global optimum of satisfaction in non-substitutive relay models.]{
\label{contobest} 
\includegraphics[width=2.3in]{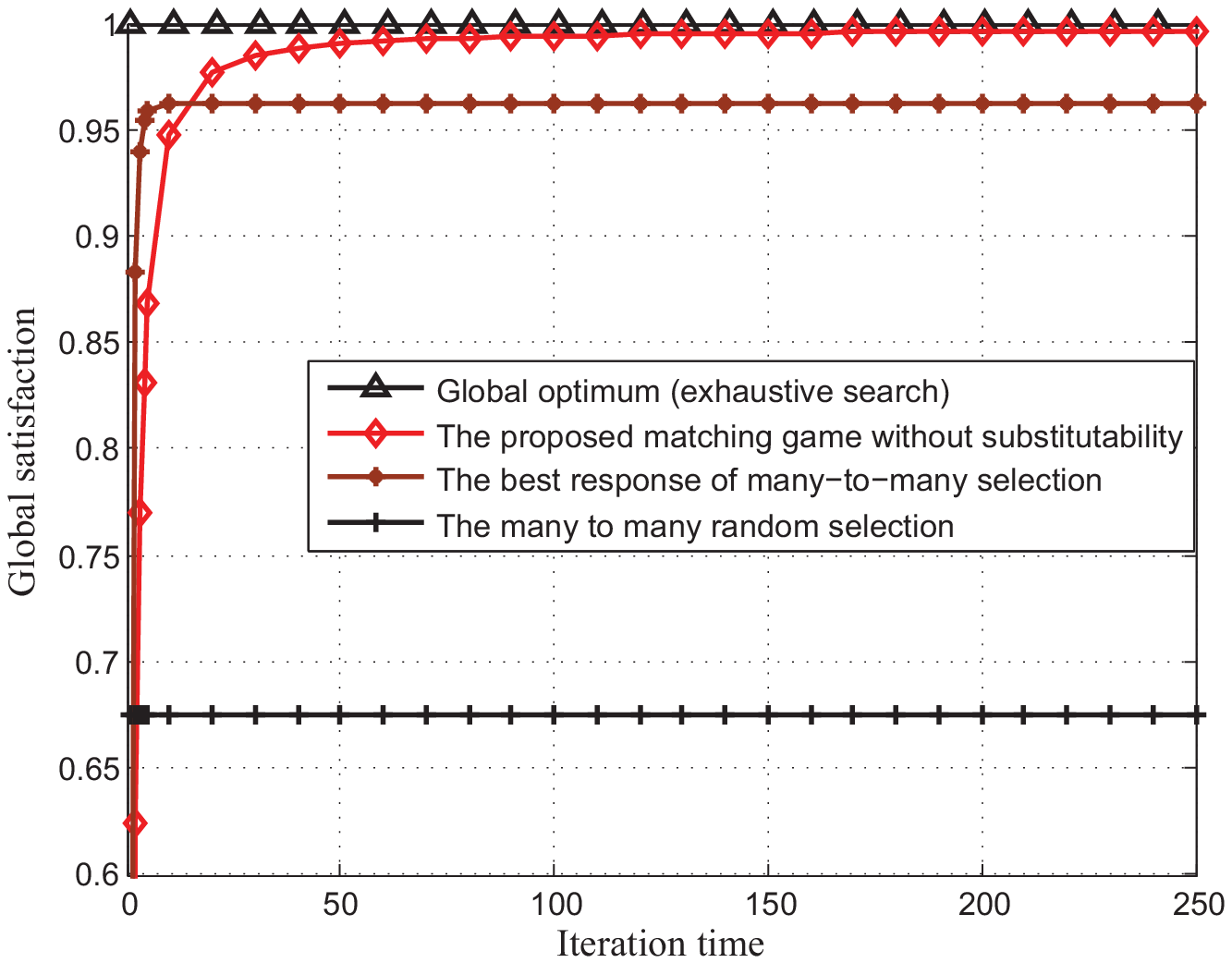}}
\centering
\subfigure[Dynamics of satisfaction performance with perturbations. At iteration t = 15 and 30, 8 source drones leave the system and 5 new source drones enter the system, respectively.]{
\label{dynamic2} 
\includegraphics[width=2.3in]{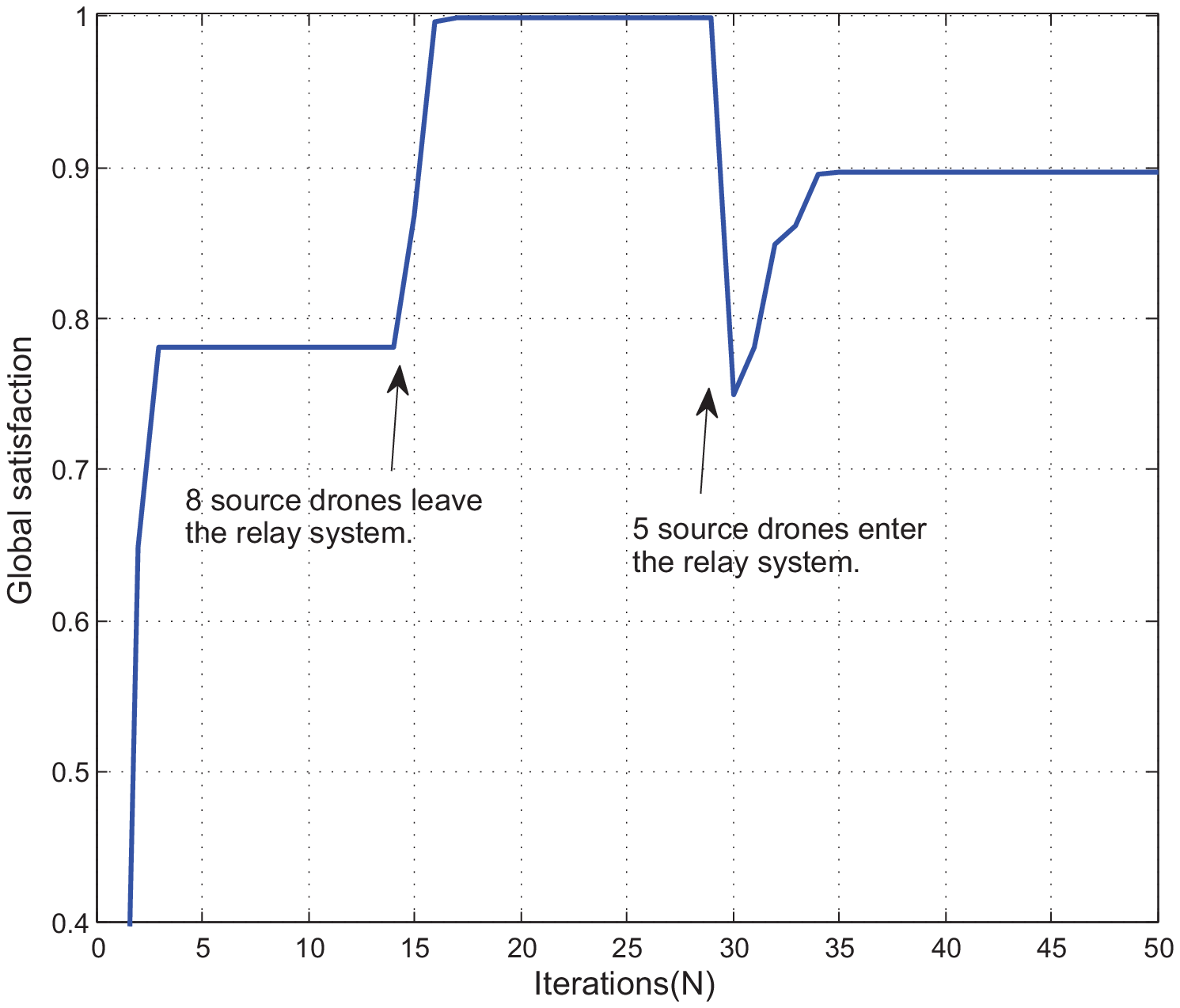}}
\caption{Simulation results of preliminary applications of the proposed classification in relay models.}
\label{fig:subfig} 
\end{figure*}

\begin{itemize}\vspace{-0.1cm}
\item \emph{Class I: Matching with substitutability}: This class constitutes the popular and baseline class, in which the relationship between players is solely competitive. Players replace other competitors when they are successfully accepted by players in the opposing set. Existing literature of matching game in wireless networks always modeled the network as the matching market with substitutability. The most prominent example is the resource allocation model with fixed matching quotas such as base station association with fixed number of channels.

\item \emph{Class II: Matching with partial substitutability}: In the second class, competitive relationships exist among matching applicants. Receiving matching requests, players filter the applicants according to the priority. However, priority is not the only filtering criterion for refusal or acceptance. The matching with partial substitutability considers the internal factors of competition such as resource surplus or tolerance limit of resource demands. The most prominent example is the resource allocation model with unfixed matching quotas, such as channel association considering unfixed time resource.

\item \emph{Class III: Matching without substitutability}: Matching market without substitutability \cite{without-Substitutability} can be used to model future networks with resource sharing. This matching model has been used in social networks. For example, one laboratory needs different types of professors to cooperatively complete research tasks.

    ~~In this class, the priority of players is not the criterion for filtering. Players do not accept or reject a matching connection according to the unilateral performance of applicants. The performance of the whole network or the applicants with other connections can be considered in the matching process. This class is suitable for the fully sharing networks, such as spectrum sharing equally with unfixed quotas.

\end{itemize}

%
%
%

\subsection{Preliminary applications of the proposed classification in relay networks}

\subsubsection{Classic matching with substitutability}
It is the classic matching model which is well studied in the existing researches. Authors such as in \cite{futurematching,newmatchinggame,matchinggame} verified that the communication systems with fixed resource blocks can achieve the stable results efficiently by the matching schemes.

\subsubsection{Multi-users selection with partial substitutability}
In \cite{multione}, we studied the problem of relay selection with multi-nodes using a many-to-one matching model.

The source nodes' preferences capture the data rates while the relay nodes filter source nodes according to the transmission efficiency and the residual resource. Due to the uncertain quota and the heterogeneous requirements, some source nodes with high data rates can not replace those with low data rates in the matching process. The reason is that the chosen source nodes are more complementary for the resource of relay nodes. It is illustrated in Fig. \ref{performancecompared10relays1000times} that, the proposed distributed matching approach based on the matching game with partial substitutability has a significant advantage in terms of satisfaction for all network sizes.


\subsubsection{Many-to-many matching networks without substitutability}
In \cite{manytomany}, we consider a UAV communication network with multi-drones equipped with multi-access interface. $20$ source drones sharing 10 relay radios of $5$ relay drones. Choosing the same channel resource, source drones will share the time resource of relay drones equally. The chosen relay drones take not only their own performance but also the matching results of other relay drones into consideration. In this case, the relay model of UAV communication networks can be modeled as a matching market without substitutability.

The average convergence performance of the distributed relay selection model is shown in Fig. \ref{contobest}. The solution based on the matching game without substitutability catches up with the global optimum. It can be noted that matching game without substitutability is better than other relay selection solutions such as best response solution, which validates the performance of matching game in the model without substitutability.

The robustness of the matching algorithm is also investigated. At iteration $t =15$ and $30$, we let 8 source drones leave the model and 5 new source drones enter the model, respectively. The results in Fig. \ref{dynamic2} show that the system can quickly converge to a stable result after the perturbations occur. This verifies that the matching game algorithm is robust to the dynamics of communication tasks in the UAV networks.


\subsection{Future research direction}
It can be seen that the matching game for self-organizing relay selection optimization in UAV communication networks has definitely drawn an exciting future, while current researches are still far away from the expected vision. We list some future research problems for matching game models and relay selection in UAV communication networks below:
\subsubsection{Diversified communications in UAV communication networks}
It can be noted that multiple transmission models exist in large UAV communication networks. Drones will join with different transmission tasks at the same time. For example, one drone connected to the coalition leader may be chosen as a relay node to assist another transmission pairs. In order to achieve better performance of UAV communication networks, some new features of game models should be developed to improve the applicability in resource allocation.
\subsubsection{Matching game with imperfect information}
Knowledge can be viewed as the high-level intelligence obtained from the contextual information, which is truly beneficial to decision-making. In most existing studies, it is assumed that all players can obtain correct information, while the information acquisition may be incomplete or even incorrect in complex systems. Such imperfect information brings about new challenges since players make selection decisions depending on the knowledge. Optimizing matching performance with imperfect information is useful for the future UAV communication networks.
\subsubsection{Design and analysis of the heterogeneous matching market}
Most existing studies assumed that players in one side of the matching market employed the same utility objective. However, the assumption is not practical in reality. In practice, drones in the network belong to different UAV coalitions, which have different service requirements or types. In addition, the same type of drones may have heterogeneous matching utilities due to different processing abilities. Introducing heterogeneity into the matching market will enhance the flexibility in optimization, which needs to be further studied.

\section{Conclusion}
This article provided a distributed matching model perspective for relay selection in UAV communication networks. First, this article introduced different situations of relay selections in large UAV communication networks, and pointed out the challenges and requirements of relay models. Then, the fundamental concepts of the matching theory and the advantages in UAV communication networks were presented. In order to understand the selection issues of relay models, this article showed the extensive applications of the matching game in wireless networks. The preliminary simulations validated the performances of the relay selection strategies. Finally, research directions of the matching model in UAV communication networks were discussed.


\begin{thebibliography}{1}

\bibitem{R2.6}
A. Adrian, C. H. Parker, and K. Tumer, ``Evolving large scale UAV communication system," \emph{in Proc. GECCO}, pp.1023-1030, 2012.

\bibitem{UAVcommunication}
L. Gupta, R. Jain and G. Vaszkun, ``Survey of important issues in UAV communication networks," \emph{IEEE Commun. Surveys Tuts.}, vol. 18, no. 2, pp. 1123-1152, 2016.

\bibitem{R2.11}
W. Qi, W. Hou, L. Guo,, et al., ``A unified routing framework for integrated space/air information networks," \emph{IEEE Access}, vol. 4, pp. 7084-7103, 2016.

\bibitem{R2.12}
I. Bekmezci, O. K. Sahingoz, and S. Temel, ``Flying ad-hoc networks (FANETs): A survey," \emph{Ad Hoc Netw.}, vol. 11, no. 3, pp. 1254-1270, 2013.

\bibitem{matchinggame}
A. Roth and M. A. O. Sotomayor, ``Two-sided matching: A study in game-theoretic modeling and analysis," \emph{Cambridge University Press}, 1992.

\bibitem{newmatchinggame}
S. Bayat, Y. Li, L. Song, et al., ``Matching theory: Applications in wireless communications," \emph{IEEE Signal Process. Mag.}, vol. 33, no. 6, pp. 103-122, 2016.

\bibitem{futurematching}
Y. Gu, W. Saad, M. Bennis, et al., ``Matching theory for future wireless networks: Fundamentals and applications," \emph{IEEE Commun. Mag.}, vol. 53, pp. 52-59, 2015.

\bibitem{R2.14}
Y. Altshuler, A. Pentland, S. Bekhor, et al., ``Optimal dynamic coverage infrastructure for large-scale fleets of reconnaissance UAVs," \emph{Springer}, 2018.

\bibitem{R2.1}
S. Sekander, H. Tabassum and E. Hossain, ``Multi-tier drone architecture for 5G/B5G cellular networks: Challenges, trends, and prospects," \emph{IEEE Commun. Mag.}, vol. 56, no. 3, pp. 96-103, Mar. 2018.

\bibitem{R2.13}
Y. Zeng and R. Zhang, ``Energy-efficient UAV communication with trajectory optimization," \emph{IEEE Trans. Wireless Commun.}, vol. 16, no. 6, pp. 3747-3760, Jun., 2017.

\bibitem{multione}
D. Liu, Y. Xu, Y. Xu, et al., ``Distributed satisfaction-aware relay assignment: A novel matching-game approach," \emph{Trans. Emerg. Telecommun. Technol.}, vol. 27, no. 8, pp. 1087-1096, 2016.

\bibitem{R2.9}
S. M. A. Kazmi, N. H. Tran, T. M. Ho et al., ``Hierarchical matching game for service selection and resource purchasing in wireless network virtualization," \emph{IEEE Commun. Lett.}, vol. 22, no. 1, pp. 121-124, Jan. 2018.


\bibitem{R1.27}
Y. Zeng, R. Zhang and T. J. Lim, ``Wireless communications with unmanned aerial vehicles: Opportunities and challenges," \emph{IEEE Commun. Mag.}, vol. 54, no. 5, pp. 36-42, May 2016.


\bibitem{without-Substitutability}
M. Pycia, ``Many-to-one matching without substitutability," \emph{MIT-IPC Working Paper}, 2005.



\bibitem{manytomany}
D. Liu, Y. Xu, Y. Xu, et al., ``Distributed relay selection for heterogeneous UAV communication networks using a many-to-many matching game without substitutability," in \emph{in Proc. IEEE/CIC ICCC}, pp. 1-6, Oct. 2017.


\end{thebibliography}
\end{document}